\documentstyle[epsf,epsfig,wrapfig,here,12pt]{article}    
\begin{document}           
\baselineskip=0.33333in
\begin{quote} \raggedleft TAUP 2845-2006
\end{quote}
\vglue 0.5in
\begin{center}{\bf The Physical Meaning of Gauge Transformations \\
in Electrodynamics}
\end{center}
\begin{center}E. Comay$^*$
\end{center}

\begin{center}
School of Physics and Astronomy \\
Raymond and Beverly Sackler Faculty of Exact Sciences \\
Tel Aviv University \\
Tel Aviv 69978 \\
Israel
\end{center}
\vglue 0.5in
\vglue 0.5in
\noindent
\vglue 0.2in
\noindent
Abstract:

The structure of classical
electrodynamics based on the variational principle
together with causality and space-time homogeneity is analyzed. It
is proved that in this case the 4-potentials are defined uniquely.
On the other hand, the approach where Maxwell equations and the Lorentz
law of force are regarded as cornerstones of the theory allows gauge
transformations. For this reason, the two theories are
{\em not equivalent}.
A simple example substantiates this conclusion.
Quantum physics is linked to the variational principle and it is
proved that the same result holds for it. The compatibility of this
result with the gauge invariance of the Lagrangian density is explained.

\newpage
\noindent
{\bf 1. Introduction}
\vglue 0.33333in

One may regard the equations of motion of a physical system as the
fundamental elements of a theory. Thus, the equations of motion can
be used for deriving useful formulas that describe properties of
the system. However, it is now recognized that other principles
play a more profound role. Using this approach, the variational
principle, causality and homogeneity of space-time are regarded
here as the basis for the discussion. The present work examines
these approaches within the validity domains of classical
electrodynamics and of quantum physics.
Thus, the electrodynamic theory that regards
Maxwell equations and the Lorentz law of force as cornerstones of
the theory is called here Maxwell-Lorentz electrodynamics (MLE).
The theory that relies on the variational principle is called here
canonical electrodynamics (CE). MLE and CE are very closely related
theories. Thus, Maxwell
equations and the Lorentz law of force can be derived from the
variational principle (see [1], pp. 49-51,70,71,78-80;
[2], 572-578,595-597).
The first part of the
discussion carried out here analyzes the two approaches
within the realm of classical electrodynamics and proves that
MLE is {\em not equivalent} to CE and that CE imposes
further restrictions on the theory's structure. Quantum mechanics is
strongly linked to the variational approach (see [3], pp. 2-23).
Thus, it is proved that the same results are valid
for quantum mechanics.

It is proved in this work that if one adheres to CE
together with causality and space-time homogeneity
then the 4-potentials of electrodynamics
are defined uniquely. On the other hand, the 4-potentials play no
explicit role
in Maxwell equations and in the Lorentz law of force. Hence, one may
apply any gauge transformation without affecting MLE.
This is the underlying reason for the claim that
MLE is {\em not equivalent} to CE.

   In the present work, units where the speed of light
$c = 1$ and $\hbar = 1$ are used. Thus, one kind of dimension
exists and the length $[L]$ is used for this purpose.
Greek indices run from 0 to 3. The metric is diagonal and its entries are
(1,-1,-1,-1). The symbol $_{,\mu }$ denotes the partial differentiation
with respect to $x^\mu $. $A_\mu $ denotes the 4-potentials and $F^{\mu \nu }$
denotes the antisymmetric tensor of the electromagnetic fields
\begin{equation}
F^{\mu \nu } = g^{\mu \alpha}g^{\nu \beta}(A_{\beta ,\alpha} -
A_{\alpha ,\beta})
=\left(
\begin{array}{cccc}
0   & -E_x & -E_y & -E_z \\

E_x &  0   & -B_z &  B_y \\

E_y &  B_z &   0  & -B_x \\

E_z & -B_y &  B_x &  0
\end{array}
\right).
\label{eq:FMUNU}
\end{equation}

In the second Section, the main point of this work is proved for classical
physics. The third Section describes a specific example that substantiates
the proof included in Section 2. The fourth Section proves that the same
results are obtain for quantum physics. The last Section contains
concluding remarks.

\vglue 0.66666in
\noindent
{\bf 2. Gauge Transformations and Canonical Electrodynamics}
\vglue 0.33333in

The Lagrangian density used for a derivation of Maxwell equations is
(see [1], pp. 78-80; [2], pp. 596-597)
\begin{equation}
{\mathcal L} =
- \frac {1}{16\pi }F^{\mu \nu }F_{\mu \nu } - j^\mu A_\mu .
\label{eq:LAGR}
\end{equation}
The following analysis examines a closed system of charges and fields.
For the simplicity of the discussion, let us examine the fields
associated with one charged particle $e$ whose motion is given.
This approach can be justified because, due to
the linearity of Maxwell equations, one finds that the fields of
a closed system of charges is a superposition of the fields of each
individual charge belonging to the system.
Let us examine the electromagnetic fields at a given space-time point
$x^\mu $. Using Maxwell equation
and the principle of causality, one can derive the retarded
Lienard-Wiechert 4-potentials (see [1], pp. 173-174; [2], pp. 654-656)
\begin{equation}
A_\mu = e\frac {v_\mu }{R^\alpha v_\alpha }.
\label{eq:LWPOTENTIAL}
\end{equation}
Here $v_\mu $ is the charge's 4-velocity at the retarded time and
$R^\mu $ is the 4-vector from the retarded space-time point to the
field point $x^\mu $. These 4-potentials defines the fields uniquely.

A gauge transformation of $(\!\!~\ref{eq:LWPOTENTIAL})$
is (see [1], pp. 52-53; [2], pp. 220-223)
\begin{equation}
A'_\mu = A_\mu - \Phi {,_\mu} .
\label{eq:GAUGE}
\end{equation}
In the following lines, the laws of CE are used in an investigation of
the form of the gauge function $\Phi (x^\mu)$.

Relying on the variational principle, one finds constraints on
terms of the Lagrangian density. Thus, the action is a Lorentz
scalar and in the unit system used here where $\hbar=1$,
it is dimensionless.
In particular, the 4-potentials $A_\mu $ must be entries of a
4-vector whose dimension is $[L^{-1}]$. This requirement is
satisfied by the Lienard-Wiechert 4-potentials $(\!\!~\ref{eq:LWPOTENTIAL})$.
Thus, also $\Phi _{,\mu }$ of $(\!\!~\ref{eq:GAUGE})$
is a 4-vector and $\Phi $ must be a
dimensionless Lorentz scalar function of the space-time coordinates.

Now, the coordinates are entries of a 4-vector. Therefore,
a Lorentz scalar depending on the space-time coordinates must be a
function of scalar variables of the following form
\begin{equation}
f_{a,b}(x^\mu ) = (x^\mu - x_a^\mu)(x_\mu - x_{b\mu}),
\label{eq:SCALAR2}
\end{equation}
where $x_a^\mu $ and $x_b^\mu $
denote specific space-time points. Relying on
the homogeneity of space-time, one finds that in the case discussed
here there is just one specific point $x_a^\mu$, which is the
retarded position of the charge. Thus, $(\!\!~\ref{eq:SCALAR2})$
is cast into the following form
\begin{equation}
f_{a,b}(x^\mu ) \rightarrow R^\mu R_\mu .
\label{eq:R2}
\end{equation}
This outcome proves that the gauge function, which is a dimensionless
quantity, must be a constant.

These arguments complete the proof showing that if one adheres to CE then
the gauge function $\Phi $
is a constant and the gauge 4-vector $\Phi _{,\mu }$ vanishes identically.
Hence, the Lienard-Wiechert 4-vector $(\!\!~\ref{eq:LWPOTENTIAL})$ is
unique.

\vglue 0.66666in
\noindent
{\bf 3. An Example}
\vglue 0.33333in

   Let us examine a simple system which consists of one motionless
particle whose mass and charge are $m$, $e$, respectively. The particle
is located in a spatial region
where the external fields vanish. Therefore, the
Lorentz force exerted on the particle vanishes too and it
remains motionless as long as these conditions do not change. Hence,
from the point of view of MLE, the particle's energy is a constant
\begin{equation}
E = m.
\label{eq:EEQM}
\end{equation}

   Now, let us examine this system from the point of view of CE. For
this purpose, the external 4-potentials should be defined. Thus, the
null external fields are derived from null 4-potentials
\begin{equation}
A_{(ext)\mu } = 0\;\rightarrow \; F^{\mu \nu}_{(ext)}=0.
\label{eq:A0}
\end{equation}

   In order to define the particle's energy
one must construct the Hamiltonian. Here
the general expression is (see [1], pp. 47-49; [2], pp. 575)
\begin{equation}
H=[m^2 + ({\bf P} - e{\bf A})^2]^{1/2} + e\phi ,
\label{eq:HAM}
\end{equation}
where ${\bf P}$ denotes the canonical momentum and the components of
the 4-potentials are $(\phi,{\bf A})$. Substituting the null values of
$(\!\!~\ref{eq:A0})$ into $(\!\!~\ref{eq:HAM})$ and putting there
${\bf P} = 0$ for the motionless particle, one equates the energy
to the Hamiltonian's value and obtains
\begin{equation}
E = m.
\label{eq:EHAMM}
\end{equation}
At this point, one finds that result $(\!\!~\ref{eq:EEQM})$ of MLE is
identical to $(\!\!~\ref{eq:EHAMM})$ of CE.

   Now, let us apply a gauge transformation to the null external
4-potentials $(\!\!~\ref{eq:A0})$. The gauge function and its 4-potentials
are
\begin{equation}
\Phi = t^2 \;\rightarrow \; A'_{(ext)\mu} = -\Phi _{,\mu} = (-2t,0,0,0).
\label{eq:GT}
\end{equation}
In MLE nothing changes, because the equations of motion depend on
electromagnetic fields and their null value does not change
\begin{equation}
F'^{\mu \nu} = F^{\mu \nu} = 0.
\label{eq:FEQF}
\end{equation}
Hence, the energy value $(\!\!~\ref{eq:EEQM})$ continues to hold and
the gauge transformation $(\!\!~\ref{eq:GT})$ is acceptable in MLE.

   The following points show several arguments proving that
this conclusion does not hold for the CE theory.
\begin{itemize}
\item[{1.}] The gauge function of $(\!\!~\ref{eq:GT})$ has the dimensions
$[L^2]$, whereas in CE it must be dimensionless.
\item[{2.}] The gauge function of $(\!\!~\ref{eq:GT})$ is the entry $U^{00}$
of the second rank tensor $U^{\mu \nu } = x^\mu x^\nu $. On the other
hand, in CE the gauge function must be a Lorentz scalar.
\item[{3.}] Substituting the gauge 4-vector $A'_{(ext)\mu }$ of
$(\!\!~\ref{eq:GT})$ into the Hamiltonian $(\!\!~\ref{eq:HAM})$, one finds
the following value for the energy
\begin{equation}
E' = H' = m - 2et.
\label{eq:HGT}
\end{equation}
Hence, if gauge transformations are allowed in CE then the energy of a
closed system is not a constant of the motion.
\end{itemize}
These three conclusions prove that a gauge transformation destroys
CE.

\vglue 0.66666in
\noindent
{\bf 4. Gauge Transformations and Quantum Physics}
\vglue 0.33333in

As stated in the Introduction, quantum physics is very closely related
to CE. Moreover, the Ehrenfest theorem (see [5], pp. 25-27, 138) shows
that the classical limit of quantum mechanics agrees with the laws of
classical physics. For these reasons, one expects that the laws of CE
are relevant to quantum physics. A direct examination of gauge
transformations proves this matter.

The Lagrangian density of the Dirac field is (see [3], p. 84; [4],
p. 78)
\begin{equation}
{\mathcal L} = \bar \psi[\gamma ^\mu (i\partial _\mu - eA_\mu) - m]\psi ,
\label{eq:DIRACLD}
\end{equation}
Now, in quantum mechanics, the gauge transformation $(\!\!~\ref{eq:GAUGE})$
is accompanied by an appropriate transformation of the particle's wave
function. Thus, the quantum mechanical form of gauge transformation is
(see [4], p. 78)
\begin{equation}
A'_\mu = A_\mu - \Phi {,_\mu} ;\;\;\;
\psi '(x^\mu ) = e^{ie\Phi(x^\mu)}\psi (x^\mu )
\label{eq:GAUGEQM}
\end{equation}
(Note that the symbol $e$ in the exponent denotes the particle's electric
charge.)
Substituting the gauge transformation $(\!\!~\ref{eq:GAUGEQM})$
into the Lagrangian density $(\!\!~\ref{eq:DIRACLD})$, one realized
the it is gauge invariant indeed (see e.g. [4], p. 78).

   Now let us examine the quantum mechanical version of the example
discussed in Section 3. The Dirac wave function of the spin-up
state of a motionless particle is (see [6], p. 10)
\begin{equation}
\psi (x^\mu ) = e^{-imt}(1,0,0,0).
\label{eq:PSI0}
\end{equation}
Thus, one uses the fundamental quantum mechanical equation and obtains
the particle's energy from an application of the
Dirac Hamiltonian to the wave function $(\!\!~\ref{eq:PSI0})$
\begin{equation}
E\psi = H\psi =
i\frac {\partial \psi }{\partial t}  = m\psi \rightarrow E=m.
\label{eq:EQM}
\end{equation}

Now, let us examine the gauge transformation $(\!\!~\ref{eq:GAUGEQM})$ for
the specific case $(\!\!~\ref{eq:GT})$. The
wave function $(\!\!~\ref{eq:PSI0})$ transforms as follows
\begin{equation}
\psi '(x^\mu ) = e^{iet^2}e^{-imt}(1,0,0,0).
\label{eq:PSITAG}
\end{equation}
A straightforward
calculation of the energy for the gauge transformed wave function
$(\!\!~\ref{eq:PSITAG})$ proves that the result differs from the original
value
\begin{equation}
E'\psi '= i\frac {\partial \psi '}{\partial t}  = (m - 2et)\psi '
\rightarrow E' = m - 2et.
\label{eq:EQMTAG}
\end{equation}

This is precisely the same discrepancy which was found above for the
gauge transformation of CE of classical physics $(\!\!~\ref{eq:HGT})$.
Thus, one concludes that gauge transformations are
inconsistent with quantum physics.

\vglue 0.66666in
\noindent
{\bf 5. Conclusions}
\vglue 0.33333in

The foregoing results indicate the difference between an electrodynamic
theory where Maxwell equations and the Lorentz law of force are regarded
as the theory's cornerstones and a theory based on the variational
principle together with
causality and space-time homogeneity. Indeed, if Maxwell
equations and the Lorentz law of force are the theory's cornerstone then
it is very well known that one is free to
define the gauge function $\Phi(x^\mu )$ of $(\!\!~\ref{eq:GAUGE})$
(see [1], pp. 52-53; [2], pp. 220-223). On the other hand,
this work proves that gauge transformations are inconsistent with
electrodynamics based on the variational principle.
For this reason, one concludes
that the two approaches are {\em not equivalent.} It is also proved
that gauge transformations are forbidden in quantum physics. 

The outcome of this work does not negate the well known gauge invariance
of the Lagrangian density. Indeed, in the Dirac Lagrangian density
$(\!\!~\ref{eq:DIRACLD})$, the two parts of the gauge transformation
$(\!\!~\ref{eq:GAUGEQM})$ cancel each other. On the other hand, the
Dirac Hamiltonian contains only one term of $(\!\!~\ref{eq:GAUGEQM})$.

\newpage
References:
\begin{itemize}
\item[{*}] Email: elic@tauphy.tau.ac.il  \\
\hspace{0.5cm}
           Internet site: http://www-nuclear.tau.ac.il/$\sim $elic
\item[{[1]}] L. D. Landau and E. M. Lifshitz, {\em The Classical
Theory of Fields} (Elsevier, Amsterdam, 2005).
\item[{[2]}] J. D. Jackson, {\em Classical Electrodynamics} (John Wiley,
New York,1975).
\item[{[3]}] J. D. Bjorken and S.D. Drell, {\em Relativistic Quantum
Fields} (McGraw-Hill, New York, 1965).
\item[{[4]}] M. E. Peskin and D. V. Schroeder, {\em An Introduction to
Quantum Field Theory} (Addison-Wesley, Reading, Mass., 1995).
\item[{[5]}] L. I. Schiff, {\em Quantum Mechanics} (McGraw-Hill, New York,
1955).
\item[{[6]}] J. D. Bjorken and S.D. Drell, {\em Relativistic Quantum
Mechanics} (McGraw-Hill, New York, 1964).

\end{itemize}

\end{document}